\documentclass[nohyper,notoc]{JHEP} 

\usepackage{epsfig}
\usepackage{epsf}

\newcommand\fverb{\setbox\pippobox=\hbox\bgroup\verb}
\newcommand\fverbdo{\egroup\medskip\noindent%
                        \fbox{\unhbox\pippobox}\ }
\newcommand\fverbit{\egroup\item[\fbox{\unhbox\pippobox}]}
\newbox\pippobox
\newcommand{\beq}{\begin{equation}}
\newcommand{\eeq}{\end{equation}}
\newcommand{\beqa}{\begin{eqnarray}}
\newcommand{\eeqa}{\end{eqnarray}}

\newcommand{\krig}[1]{\stackrel{\circ}{#1}}
\newcommand{\barr}[1]{\not\mathrel #1}
\title{On wave function renormalization and related aspects
in heavy fermion effective field theories}

\author{Sven Steininger\thanks{Work supported in part by funds
    provided by the Graduiertenkolleg ``Die Erforschung subnuklearer
    Strukturen der Materie'' at Bonn University and by DAAD.}
 \\
        Physics Department,  Harvard University,
        Cambridge MA 02138, USA\\  {\rm and}\\
        Forschungszentrum J\"ulich, Institut f\"ur Kernphysik (Theorie)\\
        D-52425 J\"ulich, Germany\\ 
        E-mail: \email{S.Steininger@fz-juelich.de}}
\author{Ulf-G. Mei{\ss}ner, Nadia Fettes\\
        Forschungszentrum J\"ulich, Institut f\"ur Kernphysik (Theorie)\\
        D-52425 J\"ulich, Germany\\
        E-mail: \email{Ulf-G.Meissner@fz-juelich.de},
                \email{N.Fettes@fz-juelich.de}}
\received{\today}               
\accepted{\today}               

\preprint{FZJ-IKP(TH)-1998-21}  

\abstract{
We reconsider the question of wave function renormalization in heavy
fermion effective field theories. In particular, we work out a simple
and efficient scheme to define the wave function renormalization with respect
to the lowest order heavy fermion propagator. The method presented is free
of a set of ambiguities which arise in heavy fermion effective field theories.
In this context, we discuss the approaches used in the literature so
far. We also calculate the fourth order pion mass contribution to the
nucleon mass shift and discuss the tree and loop contributions to the 
electric Sachs form factor of the nucleon.
}


\begin{document}


\section{Introduction}

Quantum Chromodynamics (QCD) admits two interesting limits, which can be
treated by similar methods. In the sector of the heavy quarks ($c$
and $b$) one observes to leading order the so--called heavy quark
and the related spin symmetry~\cite{HQ}. The heavy quarks can be
treated in a non--relativistic framework which  is called Heavy
Quark Effective Field Theory (HQEFT), reminiscent of the well-known
Foldy--Wouthuysen transformation for heavy Dirac particles. The
QCD Lagrangian for the heavy flavors, collectively denoted by the
field $Q$, takes the simple form
\beq
{\cal L}_{\rm QCD} = \bar{Q} (i \barr{D} - M_Q) Q~,
\eeq
where all the propagation of the heavy quarks and their
interactions can be handled as power expansions in $1/M_Q$.
Similarly, for the light quark sector, one observes
that the current quark masses are small compared to the typical scale
of the strong interactions. This defines the chiral limit of QCD,
which can be analysed making use of chiral perturbation
theory (CHPT)~\cite{gl85}. This is the effective field theory of the Standard
Model at low energies, with the pseudo--Goldstone bosons $\pi$, $K$,
$\eta$ the pertinent degrees of freedom.  In the presence of matter
fields, like e.g. baryons, a complication arises due to the large
baryon mass scale, which is comparable to the scale of chiral symmetry
breaking, $m_B \sim \Lambda_\chi \sim 1\,$GeV.
 Resorting to methods taken from HQEFT, Jenkins and
Manohar~\cite{jm} showed how to move the troublesome baryon mass term
in a string of $1/m_B$ suppressed interaction vertices. This allows
for a consistent power counting since the baryon propagator to leading
order is given by
\beq\label{prop}
S(\omega) = \frac{i}{\omega + i\varepsilon} \,\, \quad
(\varepsilon \to 0^+)\,\,\, ,
\eeq
with $\omega = v\cdot k$ in terms of the four--velocity vector
$v_\mu$ and the small residual momentum $k$, $v\cdot k \ll m_B$.
As wanted, the baryon mass has disappeared from the propagator.
The simplest way to systematically calculate the $1/m_B$ corrections
was spelled out in ref.\cite{bkkm}, in which the path integral formalism
for HQEFT developed by Mannel et al.~\cite{MRR} was extended to baryon CHPT.

Clearly, in heavy baryon CHPT as well as in HQEFT, the massive
degrees of freedom behave essentially non--relativistically and
it is thus not obvious how to extend the notion of wave function
renormalization to such a situation. In relativistic baryon CHPT,
this is not an issue since one can apply standard quantum field
theoretical methods, as detailed in~\cite{gl85}\cite{gss}. In the work of
ref.~\cite{bkkm}, the wave function renormalization was defined
via the derivative of the nucleon self--energy at $\omega = 0$, leading to a
momentum independent result for $Z_N$, the heavy nucleon Z--factor.
Since the propagator eq.(\ref{prop}) develops a pole at
this value of $\omega$, it  is the natural point to expand around, but
one could
equally chose other values of $\omega$ to define the Z--factor.\footnote{We
are much indebted to Thomas Hemmert for clarification on this topic.}
A somewhat different interpretation was given in ref.\cite{fear}.
A more detailed analysis of this particular aspect was performed by
Ecker and Moj\v zi\v s~\cite{eckmojwf}, who argued that the Z--factor
can not be a constant but rather depends (in momentum space) on the
chosen frame via the baryon momentum. They for the first time
stressed the role of the heavy fermionic sources and within their
scheme, the contribution from these sources is entirely given by
$Z_N$ and one thus does not have to perform any explicit calculation
for terms involving these heavy sources (once the Z--factor is
determined). Note that the Z--factor given in that paper
for the ``BKKM'' approach is not correct, it should be
momentum--independent. Such an observation was independently made in
ref.\cite{bimg}.\footnote{We are grateful to Gerhard Ecker for
  confirmation on this statement.}
This momentum dependence is,
however, also present in the treatment {\`a} la ref.\cite{bkkm}.
In that approach, the tree graphs are calculated from the
{\it relativistic} tree level Lagrangian and then expanded in inverse powers
of the nucleon mass up to the needed accuracy. More precisely, one
has to include all relativistic tree level Lagrangian terms
${\cal L}_{\pi N}^{(1)} + {\cal L}_{\pi N}^{(2)} + \ldots$ which, when
expanded, can  contribute
to the order of $M_\pi /m_N$ one is after.  This procedure
contains automatically the momentum dependent pieces through the
spinor normalization, as explicitely shown for the case of elastic
pion--nucleon scattering in~\cite{FMS}. A general proof that this
method always leads to the correct results has not been given. The
aim of this paper is to set up a very simple scheme for wave
function renormalization in heavy fermion effective field
theories which parallels as closely as possible the
conventional quantum field theory approach. As we will show, it is
very useful to elucidate the interrelationship between the various
approaches found in the literature. It also provides us with the
proof that expanding the relativistic tree graphs indeed leads to
the correct result, as first conjectured in~\cite{bkmpi0}. The
method developed should also be of interest for HQEFT, since to our
knowledge this issue has not been addressed in detail there. An
exception to this is ref.\cite{balk} in which the question of
how to properly normalize spinors in HQEFT is discussed.

The manuscript is organized as follows. In sec.~\ref{sec:HBCHPT} we
review the salient features of heavy baryon chiral perturbation theory
(HBCHPT) necessary to keep our presentation self--contained.
In sec.~\ref{sec:WF} we establish a novel scheme to define wave function
renormalization in heavy fermion EFTs, based on a set of four simple
conditions and interpreting the so--called light components of the
heavy fermions as Dirac spinors. This scheme is free of a set of
ambiguities, which naturally arise in heavy fermion effective field theories.
Section~\ref{sec:Pauli} is devoted to an alternative interpretation,
in that the light components  are
treated as two--component Pauli spinors. This allows us to establish the
relation between the general definiton given before and the method
of expanding the relativistic tree graphs, as commonly done~\cite{bkmrev}.
Since S--matrix elements and transition currents do, of course, not depend
on the way one defines wave function renormalization, we consider
in detail  the electric Sachs form factor in sec.~\ref{sec:charge}.
We show that all different calculational schemes lead to a constant
contribution from the Born terms in conjunction with the appropriate
wave function renormalization. We also show (in an appendix) that the loop graphs
do not renormalize the charge, as it should be. The final
comments and summary are given in sec.~\ref{sec:summ}. Some
technicalities and additional remarks are relegated to the appendices.

\section{Heavy nucleon effective field theory}
\label{sec:HBCHPT}
\def\theequation{\arabic{section}.\arabic{equation}}
\setcounter{equation}{0}

We briefly review the path--integral formulation of the chiral
effective pion--nucleon system. This follows largely the original
work of \cite{bkkm}, which was reviewed in \cite{bkmrev}.
The interactions of the pions with the nucleons
are severely constrained by chiral symmetry. The
generating functional for Green functions of quark currents
between single nucleon states, $Z[j,\eta, \bar \eta$], is defined via
\beq
\exp \, \bigl\{ i \, Z[j,\eta, \bar \eta] \bigr\}
= {\cal N} \int [du] [dN] [d{\bar N}] \, \exp
\, i \biggl[ S_{\pi\pi} + S_{\pi N} +
\int d^4x (\, \bar \eta N  + \bar N \eta \, ) \biggr] \, \, \, ,
\label{defgenfun}
\eeq
with $S_{\pi\pi}$ and $S_{\pi N}$ denoting the pion and the pion--nucleon
effective action, respectively, to be discussed below. $\eta$ and
$\bar \eta$ are fermionic sources coupled to the baryons and $j$
collectively denotes the external fields of vector ($v_\mu$),
axial--vector ($a_\mu$), scalar ($s$) and pseudoscalar ($p$)
type.\footnote{The external vector field $v_\mu$ should not be confused
with the four-velocity to be defined later on, which is denoted by the
same symbol.}  These are coupled in the standard chiral
invariant manner. In particular, the scalar source contains the quark
mass matrix ${\cal M}$, $s(x) = {\cal M} + \ldots$.
The underlying effective Lagrangian
can be decomposed into a purely mesonic ($\pi\pi$) and a pion--nucleon
($\pi N$) part as follows (we only consider processes with exactly one
nucleon in the initial and one in the final state)
\beq
{\cal L}_{\rm eff} = {\cal L}_{\pi\pi} + {\cal L}_{\pi N}~,
\eeq
subject to the following low--energy expansions
\beq
{\cal L}_{\pi\pi} =  {\cal L}_{\pi\pi}^{(2)}
 + {\cal L}_{\pi\pi}^{(4)} + \ldots  \, \, ,
\quad {\cal L}_{\pi N} =  {\cal L}_{\pi N}^{(1)} + {\cal L}_{\pi N}^{(2)}
+ {\cal L}_{\pi N}^{(3)} + {\cal L}_{\pi N}^{(4)}  \ldots \, ,
\eeq
where the superscript denotes the chiral dimension.
The pseudoscalar Goldstone fields, i.e. the pions, are collected in
the  $2 \times 2$ unimodular, unitary matrix $U(x)$,
$ U(\phi) = u^2 (\phi) = \exp \lbrace i \phi / F \rbrace$
with $F$ the pion decay constant (in the chiral limit).
The external fields appear  in the following chiral invariant
combinations:
$r_\mu = v_\mu +a_\mu \, ,$ $l_\mu = v_\mu -a_\mu \, ,$ and
$\chi = 2 B_0 \,(s+ip)$. Here, $B_0$ is related to the quark condensate
in the chiral limit, $B_0 = |\langle 0|\bar q q|0 \rangle|/F^2$.
We adhere to the standard chiral counting, i.e. $s$ and $p$ are
counted as ${\cal O}(q^2)$, with $q$ denoting a small momentum or meson mass.
The effective meson--baryon Lagrangian starts with terms of dimension one,
\beqa
{\cal L}_{\pi N}^{(1)} = \bar \Psi \, \biggl(\, i D\!\!\!\!/  - m_0
+ {\krig{g}_A \over 2} \, u\!\!\!/ \gamma_5  \, \biggr) \Psi \,\,\, ,
\label{LMB1}
\eeqa
with $m_0$ the nucleon mass in the chiral limit and $u_\mu = i [ u^\dagger
(\partial_\mu-ir_\mu) u - u (\partial_\mu-il_\mu) u^\dagger ]$.
The nucleons, i.e. the proton and the neutron, are collected in the
iso--doublet $\Psi$, $\Psi^T = (p,n)$.
Under $SU(2)_L \times SU(2)_R$, $\Psi$  transforms as any matter field.
$D_\mu$ denotes the covariant derivative,
$D_\mu  \Psi = \partial_\mu \, \Psi +  \Gamma_\mu \Psi$
and $\Gamma_\mu$ is the chiral connection,
$\Gamma_\mu = \frac{1}{2}\, [ u^\dagger (\partial_\mu-ir_\mu) u +
u (\partial_\mu-il_\mu) u^\dagger ]$. Note that the first term in
Eq.(\ref{LMB1}) is of dimension one since $( i D \!\!\!\!/ - m_0 )\,
\Psi = {\cal O}(q)$ \cite{krause}. The lowest order
pion--nucleon Lagrangian contains two parameters, namely $m_0$ and
$\krig{g}_A$. Treating the nucleons as
relativistic spin--1/2 fields, the chiral power counting is
considerably complicated due to the large mass scale $m_0$, $\partial_0 \, \Psi
\sim m_0 \, \Psi \sim \Lambda_\chi \, \Psi$, with $\Lambda_\chi \sim 1\,$GeV
the scale of chiral symmetry breaking. A detailed analysis of this
topic can be found in~\cite{gss}. This problem can be overcome in
the heavy mass formalism proposed in \cite{jm}.
We follow here the path integral approach  developed in \cite{bkkm}.
Defining velocity--dependent spin--1/2 fields by a particular choice
of Lorentz frame and decomposing the fields into their velocity
eigenstates (also called 'light' and 'heavy' fields),
\beq
H_v (x) = \exp \{ i m_0 v \cdot x \} \, P_v^+ \, \Psi (x) \, , \quad
h_v (x) = \exp \{ i m_0 v \cdot x \} \, P_v^- \, \Psi (x) \,\, ,
\label{Bheavy}
\eeq
the mass dependence is shuffled from the fermion propagator into a
string of $1/m_0$ suppressed interaction vertices. The projection operators
appearing in Eq.(\ref{Bheavy}) are given by
\beq
P_v^\pm = \frac{1 \pm v \!\!\!/}{2}~, \quad P_v^+ H = H \, , \, P_v^-
h = h \, , \quad P_v^+ + P_v^- = 1~,
\eeq
with $v_\mu$ the four--velocity subject to
the constraint $ v^2 = 1$. To be specific, the nucleon four--momentum
has the form
\beq\label{pmu}
p_\mu = m_0 \, v_\mu + k_\mu \,\,\, ,
\eeq
where $k_\mu$ is a small residual momentum, $v \cdot k \ll m_0$.
In the basis of the velocity projected light and heavy fields, the
effective pion--nucleon action takes the form
\beq\label{LABC}
S_{\pi N} = \int d^4x \, \biggl\{ \bar{H}_v \, A \, H_v
- \bar{h}_v \, C\, h_v + \bar{h}_v \, B \, H_v
+ \bar{H}_v \, \gamma_0 \,  {B}^\dagger \, \gamma_0 \, h_v
\biggr\}\,\,\, .
\eeq
The matrices $A$, $B$ and $C$ admit low energy expansions, e.g.
\beq
A = A^{(1)}+ A^{(2)} + A^{(3)} + A^{(4)} + \ldots \,\, ,
\eeq
and similarly for $B$ and $C$.
Explicit expressions for the various contributions can be found in
\cite{bkmrev}.  Furthermore, we split the
baryon source fields $\eta (x)$ into velocity eigenstates,
\beq
R_v (x)= \exp \{ i m_0 v \cdot x \} \, P_v^+ \, \eta(x) \, , \quad
\rho_v (x) = \exp \{ i m_0 v \cdot x \} \, P_v^- \, \eta(x) \,\, ,
\label{sourceheavy}
\eeq
and shift variables, $h_v = h_v - C^{-1} \, ( B \, H_v + \rho_v )$,
so that the generating functional takes the form
\beq
\exp[iZ] = {\cal N} \, \Delta_h \, \int [dU][dH_v][d\bar{H}_v] \, \exp
\bigl\{iS_{\pi\pi} + i S_{\pi N}' \, \bigr\}
\label{Zinter}
\eeq
in terms of the new pion--nucleon action $S_{\pi N}'$,
\beqa\label{Sprime}
S_{\pi N}' &=& \int d^4x \, \biggl\{ \,
\bar{H}_v \bigl( A^{} + \gamma_0
B^\dagger \gamma_0 \, C^{-1} B \, \bigr) H_v
+ \bar{H}_v \bigl( R_v + \gamma_0 B^\dagger \gamma_0
C^{-1} \rho_v \bigr) \nonumber \\
&& \qquad\qquad\qquad\qquad + \bigr( \bar{R}_v + \bar{\rho}_v
C^{-1} \, B \bigr) H_v + \bar{\rho}_v C^{-1} \, \rho_v 
\biggr\} \, \, .
\eeqa
The determinant $\Delta_h$ related to the 'heavy' components is
identical to one, i.e. the positive and negative velocity sectors are
completely separated. The generating
functional is thus entirely expressed in terms of the Goldstone bosons
and the 'light' components of the spin--1/2 fields. The action is,
however, highly non--local due to the appearance of the inverse of
the matrix $C$. To render it local, one now expands $C^{-1}$ in powers
of $1/m_0$, i.e. in terms of increasing chiral dimension.
To any finite power in $1/m_0$, one can now perform the integration of
the 'light' baryon field components $H_v$ by again completing the
square,
\beq
H_v' = H_v - T^{-1} \, \bigl( R_v + \gamma_0 \, B^\dagger
\, \gamma_0 \, C^{-1} \, \rho_v \, \bigr) \, , \quad
T = A + \gamma_0 \, B^\dagger \, \gamma_0 \, C^{-1} \, B \,\,\, .
\eeq
Notice that the second term in the expression for $T$ only starts
to contribute at chiral dimension two. Finally, we  arrive at
\beq
\exp[iZ] = {\cal N}' \, \int [dU] \, \exp \bigl\{ iS_{\pi\pi}
+ i Z_{\pi N} \, \bigr\} \,\,\, ,
\label{Zfinal}
\eeq
with ${\cal N}'$ an irrelevant normalization constant. The generating
functional has thus been reduced to the purely mesonic functional.
$Z_{\pi N}$ is given by
\beqa\label{genfu}
Z_{\pi N} = - \int d^4x &\biggl\{&
 \bar{\rho}_v \, \bigl( C^{-1} \, B
\, T^{-1} \, \gamma_0 \, B^\dagger \, \gamma_0 C^{-1} -
C^{-1} \, \bigr) \, \rho_v
\nonumber \\
&+&
 \bar{\rho}_v \, \bigl( \, C^{-1} \, B \, T^{-1}
\, \bigr) \, R_v + \bar{R}_v \, \bigl( T^{-1} \, \gamma_0 \,
B^\dagger \, \gamma_0 \, C^{-1} \, \bigr) \, \rho_v
\nonumber \\
&+&  \quad \bar{R}_v \, T^{-1} \, R_v  \quad \biggr\} \,\,\,\,\, .
\label{ZMB}
\eeqa
At this point, some remarks are in order. First, physical matrix
elements are always obtained by differentiating the generating
functional with respect to the sources $\eta$ and $\bar\eta$. The
separation into the velocity eigenstates is given by the projection
operators as defined above. As shown in ref.\cite{eckmojwf}, the
chiral dimension of the `heavy' source $\rho_v \sim P_v^- \eta$ is larger
by one order than the chiral dimension of the `light' source, $R_v \sim
P_v^+ \eta$. The effective Lagrangian can be readily deduced from this action.
For later use, we give the first two terms, following the
definitions of \cite{bkmrev},
\beqa\label{LpiN2}
{\cal L}_{\pi N}^{(1)} &=& \bar{H}_v \biggl\lbrace i v \cdot D +
\krig{g}_A \, S \cdot u \, \biggr\rbrace H_v \,\,\, , \nonumber \\
{\cal L}_{\pi N}^{(2)} &=& \bar{H}_v \biggl\lbrace
{1\over 2m_0} \, (v\cdot D)^2 - {1\over 2m_0} D^2
 - {i \krig{g}_A \over 2m_0} \, v^{\mu} \, S^\nu  \, \{D_\nu , u_\mu \}
 + c_1 \, \langle \chi_+ \rangle   \nonumber \\
&& +\Big( c_2 - {\krig{g}_A^2 \over 8m_0}  \Big)\,(v\cdot u)^2
+ c_3 \, u \cdot u +\ldots
\biggr\rbrace H_v \,\,\, ,
\eeqa
where the ellipsis stand for three other terms not needed for the
following discussions. $S^\mu$ is the covariant spin--operator {\`a} la
Pauli--Lubanski,
$S^\mu = \frac{i}{2} \, \gamma_5 \, \sigma^{\mu \nu} \, v_\nu$
subject to the constraint $S \cdot v = 0$ and we work in the isospin limit
$m_u = m_d$. Traces in flavor space are denoted by $\langle...\rangle$.
Notice that the spin--matrices appearing in the operators have all to be
taken  in the appropriate order.
The explicit symmetry
breaking is encoded in the matrices $\chi_\pm = u^\dagger \chi
u^\dagger  \pm  u  \chi^\dagger  u $. The $c_i$ are finite low--energy
constants (LECs), their values have been
determined~\cite{bkmlec}\cite{moj}\cite{FMS}.

At one loop level, divergences appear. These can be extracted either
by direct Feynman graph calculations or, more elegantly, directly
from the irreducible generating functional~\cite{ecker}\cite{guido}.
Here we give the form relevant to fourth order in SU(2), the details
can be found in~\cite{mms4}
\beqa
Z_{\rm irr}[j,R_v] &=& \int d^4x \, d^4x' \, d^4y \, d^4y' \,
\bar{R}_v (x) \, S_{(1)}^{\rm cl} (x,y) \, \bigl[ \,
\Sigma_2^{(1)+(2)}(y,y') \,  \delta(y-y') \nonumber \\
&& \qquad + \, \Sigma_1^{(1)+(2)} (y,y') + \Sigma_3^{(2)} (y,y') \, \bigr] \,
S_{(1)}^{\rm cl}(y',x') \, R_v (x')
\label{Zirr}
\eeqa
in terms of the self--energy functionals $\Sigma_{1,2,3}$,
and $S_{(1)}^{\rm cl}$ denotes the classical propagator in the
presence of external fields. $\Sigma_1$
refers to the self-energy graphs at order $q^3$ and the same diagram
with one dimension two insertion on the nucleon line. $\Sigma_2$
collects the tadpoles at orders $q^3$ and $q^4$ and  $\Sigma_3$ refers
to the dimension two vertex corrected self--energy diagrams. The whole
machinery and the complete fourth order counterterm Lagrangian is
spelled out in~\cite{mms4}.

So far, there exist three different approaches how to calculate matrix
elements.
The first one is based on ref.\cite{bkkm} and will be referred to as
``BKKM'' in what follows. It amounts to a ``hybrid'' calculation.
The tree graphs are worked out from the  relativistic
pion--nucleon Lagrangian and then expanded in powers of $1/m_N$
to the order one is interested in. For the reasons mentioned above, the
loop graphs are calculated in the heavy nucleon framework. In
particular, the light fields $H$ are treated as Pauli spinors and
the corresponding Z--factor is entirely given by the loop graphs and
is momentum--independent. This method is very convenient for
calculations and gives
the correct results to orders $q^3$ and $q^4$ as will be shown later.
The disadvantages of the method are twofold. First, such a hybrid
approach does not appeal to everybody and second, it is
not clear how it can be extended correctly to higher orders.
Second, Ecker and Moj\v zi\v s~\cite{eckmojwf} have set up
a scheme which stays entirely within the heavy fermion approach, however,
matrix elements are  matched to the corresponding relativistic ones.
This method should be applicable at any order. The derivation rests on
the
interpretation of the $H$ fields as Dirac spinors.
The tree graphs are all calculated within
HBCHPT. It is important to note
that a {\it different} Lagrangian is used as compared to
BKKM. This Lagrangian is subject to field transformations to eliminate
the equation of motion terms and can be found in~\cite{eckmoj}.
In this framework and for the Lagrangian of~\cite{eckmoj},
the wave function renormalization is
momentum--dependent (with the exception of forward matrix elements),
\beq
Z_N (Q) = 1 +\frac{4a_3 M_\pi^2}{m_N^2} + \frac{Q^2}{4m_N^2} + \ldots~,
\eeq
where the residual momentum $Q$ is defined in terms of the on--shell
nucleon momentum $p_N$ and its physical mass $m_N$, $p_N = m_N v + Q$.
The LEC $a_3$ is related to $c_1$ in eq.(\ref{LpiN2}) via $a_3 = m_N \, c_1$.
This method has the advantage of staying within one given field
theoretical framework, however, 
the Z--factor they give for the BKKM approach should be
momentum--independent.
For a more detailed discussion of this approach and the matching to
relativistic matrix elements, we refer to ref.~\cite{eckmojwf}.
Third, a variant of the BKKM approach, which can easily be extended to higher
orders, has been proposed by Fettes et al.~\cite{FMS} (called FMS from
here on). Again, the light
fields are treated as Pauli--spinors. The tree (Born) graphs are,
however, calculated in the heavy baryon limit and the Z--factor
consists of two pieces,
\beq{\label{ZFMS}}
Z_N  = {\cal N}^2 \, Z_N^{\rm loop}~,
\eeq
with ${\cal N}$ the relativistic spinor normalization,
\beq\label{eq:N}
{\cal N} = \sqrt{\frac{E_p + m_N}{2m_N}}~,
\eeq
where $E_p = \sqrt{{\vec p\,}^2 + m_N^2}$ denotes the full
relativistic nucleon energy. For the case of pion--nucleon
scattering to order ${\cal O}(q^3)$, it was demonstrated that this
method  reproduces the result from
expanding the relativistic tree graphs, i.e. the BKKM approach. In
fact, as we will show later, one can show this equivalence quite
generally for all one--loop processes including fourth order.
This method has the advantage that it also stays within HBCHPT and
is thus not a hybrid type of calculation. However, the important
normalization factor ${\cal N}^2$ which enters the Z--factor, eq.(\ref{ZFMS}),
is not directly given by the heavy baryon theory.
It is important to note that
in the rest--frame $v_\mu = (1,\vec{0}\,)$ a connection
between the relativistic and the Pauli--spinor interpretation
is given through the relation
\begin{eqnarray}\label{PD}
P_v^+\,u  = {\cal N}\, \left(
\matrix  { \chi \nonumber \\ 0 \nonumber \\}
\!\!\!\!\!\!\!\!\!\!\!\!\!\!\!\!\! \right)  \, \, \, .
\end{eqnarray}
with $u$ ($\chi$) denoting the conventional four(two)--component Dirac (Pauli)
spinor.
All three methods used
so far share one further common disadvantage. One can only calculate
the Z--factor if one approaches the physical pole in the direction of
the four--velocity $v$, i.e. when the nucleon four--momentum $p$
defined in terms of the nucleon mass in the chiral limit, $m_0$, is taken to
its on--shell value $p_N$ defined in terms of the physical mass $m_N$,
the difference is $(p_N - p)_\mu \sim v_\mu$. Clearly, it would be
preferable to have a
definition of the wave function renormalization that does not
depend on how one approaches the physical on--shell point. In
particular, in the Pauli--spinor interpretation one selects the rest--frame
as the preferred frame and that makes the $Z$--factor
depending on this particular choice of the four--velocity vector.
The aim of the present work is to set up a scheme which does not have the
various short--comings discussed so far and allows to give a very simple and
concise
definition of wave function renormalization following as closely as
possible conventional quantum field theories. Furthermore, as we will
demonstrate, in this novel scheme it is particularly easy to compare
the existing approaches and pin down the pertinent differences.
Clearly, since the Z--factor is not an observable, one is free to
chose ones own definition, in particular also the momentum, about
which one expands. The only condition to be fulfilled is to work
consistently within the chosen framework.

\section{Wave function renormalization reconsidered}
\label{sec:WF}
\def\theequation{\arabic{section}.\arabic{equation}}
\setcounter{equation}{0}

In this section, we wish to establish a scheme,  allowing for
a definition of the heavy fermion Z--factor, which we call $Z_N$ from
here on, subject to the following four conditions:
\begin{enumerate}
\item[1)] The definition of $Z_N$ should be independent of the
choice of the four--velocity vector $v$.
\item[2)] Its definition should only involve the physical fields.
\item[3)] At tree level, one should have $Z_N^{\rm tree} =1$.
\item[4)] The definition of $Z_N$ should be independent of the
way one approaches the physical on--shell momentum, $p \to p_N$.
\end{enumerate}
As for the last point, we remark that this is also stressed 
in ref.\cite{eckmojwf}, however, for the  actual calculation a
particular ``clever'' choice was made to control the on--shell
limit. While that is certainly legitimate, our aim is to
avoid such a choice from the beginning.  
Although we will establish this scheme for the particular case of
two flavor heavy baryon chiral perturbation theory, the method is
more general and can be applied to other situations. Of course, we
stress again that $Z_N$ is not an observable and that many alternative
schemes exist to define it. As we will argue, the one presented here
is particularly useful to shed light on the various calculational
approaches used so far and helps to clarify the pertinent interrelationships.
A precise definition of $Z_N$ is relegated to app.~\ref{app:Zdef}.

The starting point of our discussion is the
interpretation of the light fields $H$ as {\it Dirac} {\it spinors},
following ref.\cite{eckmojwf}. In this case,
we have to consider all fields in the generating functional. Thus,
instead of splitting the sources in light and heavy components, it is
advantageous to work with the {\it physical} external sources. We thus
slightly rewrite the generating functional, eq.(\ref{genfu}), as (the
projection operators are kept for clarity)
\begin{eqnarray}\label{genfun}
{\cal Z} = - \int d^4 x
\;\bar{\tilde{\eta}}
& \left\{ \right. &
\left.
  P^+_v (A + B' C^{-1} B)^{-1} P^+_v
+ P^-_v C^{-1} B (A + B' C^{-1} B)^{-1} P^+_v
\right. \nonumber \\
& + &
\left.
  P^+_v (A + B' C^{-1} B)^{-1} B' C^{-1} P^-_v
\right. \\
& + &
\left.
  P^-_v C^{-1} B (A + B' C^{-1} B)^{-1} B' C^{-1} P^-_v
- P^-_v C^{-1} P^-_v
\right\} \tilde{\eta}~,
\nonumber
\end{eqnarray}
with
\beq
B' := \gamma^0 \, B^\dagger \, \gamma^0~,
\eeq
and the 'tilde' reminds us that the factor $\exp \{im_0 v\cdot x\}$ is
still included in the external sources, $\tilde{\eta} = \exp \{im_0 
v\cdot x\} \, \eta$.  The physics is given by the
Green functions, i.e. the derivatives with respect to the external
sources. For S--matrix elements, we only need to concern ourselves
with the poles of the Green functions when the particles are on mass
shell. In the generating functional, eq.(\ref{genfun}), only the
term $(A + B' C^{-1} B)^{-1}$ can give rise to a pole as long as
one considers small external momenta. To specify the most general
$n$--point Green function, we define
\beq
T := A + B' C^{-1} B = T_0 + T_I \,\,\, ,
\eeq
where the index '0' indicates the absence of external fields and the
subsript '$I$' denotes the interaction matrix, which to lowest order
is nothing but the pion coupling $\sim g_A S\cdot u$. The inverse of $T$ is
given by
\beqa
T^{-1} &=& (T_0 + T_I)^{-1}
       = T_0^{-1} (1 + T_I T_0^{-1})^{-1} \nonumber \\
       &=& T_0^{-1} - T_0^{-1} \hat{T} T_0^{-1}
       = T_0^{-1} - \hat{G}_n~.
\eeqa
Note that $\hat{T}$ is nothing but the amputated amplitude for a
general process. Stated differently, it amounts to all Born graphs
with $m$ interaction vertices and $m-1$ intermediate nucleon propagators.
The pertinent $n$--point Green function  ${G}_n$ thus takes the form
\begin{eqnarray} \label{Gn}
G_n & = &
  P_v^+ \, \hat{G}_n \, P_v^+
+ P_v^- \, C_0^{-1} B_0 \, \hat{G}_n \, P_v^+
+ P_v^+ \, \hat{G}_n \, B'_0 C_0^{-1} \, P_v^-
\nonumber \\
& &
+ P_v^- \, C_0^{-1} B_0 \, \hat{G}_n \, B'_0 C_0^{-1} \, P_v^-
\nonumber \\
& = &
\left(
1 + C_0^{-1} B_0
\right)
P_v^+ \, \hat{G}_n \, P_v^+
\left(
1 + B'_0 C_0^{-1}
\right)
\nonumber \\
& = &
\left(
1 + C_0^{-1} B_0
\right) T_0^{-1} \;
P_v^+ \, \hat{T} \, P_v^+
\; T_0^{-1}
\left(
1 + B'_0 C_0^{-1}
\right)~.
\end{eqnarray}
To arrive at this result, we have used the commutation relations
between the operators $A,B$ and $C$ and the projection operators
$P^\pm_v$. We note that eq.(\ref{Gn}) agrees with eq.(11) of~\cite{eckmojwf}.
The S--matrix is now given by reinstating the external legs,
\beq
{\cal S}  =
\bar{u}(p') (\not\!p' - m_N) (1 + C^{-1}_0 B_0) T^{-1}_0
P^+_v \hat{T} P^+_v
T^{-1}_0 (1 + B'_0  C^{-1}_0) (\not\!p - m_N) u(p)~.
\eeq
Let us first calculate  $T_0^{-1}$,
\begin{eqnarray}\label{T0}
T_0 & = &
A_0 + B_0' C_0^{-1} B_0 \nonumber \\
& = &
vk + {\barr k}^\perp (2 m_0 + vk - 4 c_1 M^2)^{-1} {\barr k}^\perp
+ 4 c_1 M^2   \nonumber \\
& = &
(2 m_0 + vk - 4 c_1 M^2)^{-1}
\left(
2 m_0 vk + k^2 + 8 m_0 c_1 M^2 - 16 c_1^2 M^4
\right)  \nonumber  \\
& = &
(2 m_0 + vk - 4 c_1 M^2)^{-1}
\left(
p^2 - m_0^2 + 8 m_0 c_1 M^2 - 16 c_1^2 M^4
\right)~,
\end{eqnarray}
where $k_\mu$ is the small residual momentum and
$ k^\perp_\mu = k_\mu - (v\cdot k) v_\mu$ is the momentum orthogonal to the
direction given by the four--velocity $v$. $M^2$ is the leading term
in the quark mass expansion of the pion mass, $M_\pi^2 = M^2 [ 1 +
{\cal O}(q^2)]$.
One can factor out the term $C_0^{-1}$ since it does not have any
pole as long as one restricts oneself to small momenta. For a given
fixed order, the corrections obtained by this factorization are always
of higher order and can thus be neglected. Consequently, only the
second term in eq.(\ref{T0}) contains a pole, which leads to the
well--known mass shift
\beq
\delta m = - 4 c_1 M^2 \,\,  .
\eeq
So we are left with the calculation of
\begin{eqnarray}\label{Gnint}
(\not\!p - m_N) (1 + C^{-1}_0 B_0) T^{-1}_0
& = &
(\not\!p - m_N) (C_0 + B_0) (p^2 - m_N^2)^{-1}
\nonumber \\
& = &
(\not\!p - m_N) (\not\!p + m_N + vk (1 - \not\!v))
(p^2 - m_N^2)^{-1}
\nonumber \\
& = &
1 + 2 (\not\!p - m_N) vk (p^2 - m_N^2)^{-1} P_v^- \,\,\, .
\end{eqnarray}
Using furthermore $P_v^- P_v^+ = 0$, the S--matrix follows as
(see also app.~\ref{app:Zdef})
\beqa\label{Zone}
{\cal S} &=&
\bar{u}(p') P^+_v \hat{T} P^+_v u(p) =
\bar{u}(p') Z_N P^+_v \hat{T} P^+_v Z_N u(p)\nonumber \\
&=& \bar{u}(p')_{\rm phys}\, \sqrt{Z_N} \, P_v^+ \, \hat{T} \,
P_v^+ \, \sqrt{Z_N}\, u(p)_{\rm phys}~,
\eeqa
which means that in this case the Z--factor is exactly one, $Z_N = 1$.
Of course, there are still corrections from the loops, which will be
evaluated subsequently. We stress here that this interpretation allows
for a clear and concise definition of the Z--factor in that only the
loop graphs lead to a non--trivial contribution.

We now consider the effects of pion loops. For that, we expand around
the classical solution of the fermion propagator
in terms of pionic fluctuations~\cite{gss}.
This means for the matrices $A, B $ and $C$ defined in eq.(\ref{LABC}),
\beqa
A  &\to& A^{\rm cl} + A^{(1)} + A^{(2)} + \ldots~, \nonumber\\
B  &\to& B^{\rm cl} + B^{(1)} + B^{(2)} + \ldots~, \nonumber\\
C  &\to& C^{\rm cl} + C^{(1)} + C^{(2)} + \ldots~,
\eeqa
where the superscript '$(i)$' counts the number of pionic 
fluctuations\footnote{These
should not be confused with the chiral dimension used before.}
(for details, see e.g. refs.\cite{gss}\cite{guido}).
Note that we do
not need to consider baryonic fluctuations since we only work at small
momenta. To calculate the corresponding mass shift,
we only need to work out the influence of these fluctuations at the
pole,
\beqa\label{fluc}
T_0^{-1} & = &
[T_0^{\rm cl} + T_0^{(1)} + T_0^{(2)} + \ldots ]^{-1}
\nonumber \\
& = &
S^{\rm cl} - S^{\rm cl} T_0^{(1)} S^{\rm cl}
           - S^{\rm cl} {T_0^{(2)}} S^{\rm cl}
       + S^{\rm cl} T_0^{(1)} S^{\rm cl} T_0^{(1)} S^{\rm cl} + \ldots~,
\eeqa
where $(S^{\rm cl})^{-1} = T_0^{\rm cl}$. Integrating over the fluctuation
fields, the second term in eq.(\ref{fluc}) vanishes, whereas the third
and the fourth give the tadpole and the self--energy contribution,
respectively. We can thus bring the expression for the self--energy
into a compact form
\beqa
T_0^{-1} =
S^{\rm cl} + S^{\rm cl} \Sigma S^{\rm cl} + \ldots =
[T_0^{\rm cl} - \Sigma]^{-1}~.
\eeqa
The last equation is exact in the sense that if one were to calculate
the contributions from all irreducible graphs to $\Sigma$, the
reducible ones follow from the geometric series (as it is
well--known).  The inverse propagator follows as
\begin{eqnarray}
S^{-1}
& = &
vk + k^\perp (2 m_0 + vk - 4 c_1 M^2- 8 b_0 M^4)^{-1} k^\perp
\nonumber \\
& &
+ 4 c_1 M^2 + 8 b_0 M^4 - \hat{\Sigma}^{(3)} (vk) -\hat{\Sigma}^{(4)} (vk)
\nonumber \\
& = &
(2 m_0 + vk - 4 c_1 M^2 - 8 b_0 M^4)^{-1}
\nonumber \\
&\times &
\left\{
p^2 - m_0^2 + 2 m_0 (4 c_1 M^2 + 8 b_0 M^4) - (4 c_1 M^2 + 8 b_0 M^4)^2
\right.
\nonumber \\
& &
\left.
-(2 m_0  + vk - 4 c_1 M^2 - 8 b_0 M^4) (\hat{\Sigma}^{(3)} (vk)
+ \hat{\Sigma}^{(4)} (vk))
\right\}~,
\end{eqnarray}
with $\hat{\Sigma}^{(3,4)} (\omega )$ denoting the third and fourth order
loop contribution to the nucleons' self--energy. 
Here, $b_0$ is a combination of the LECs of three $q^4$ counter terms, which in
fact lead to a quark mass correction of the LEC $c_1$. The numerical value of 
$b_0$ is at present not known. It could be obtained from a fourth order
analysis of the baryon masses and $\sigma$--terms (for a model--dependent
determination within SU(3) baryon CHPT, see e.g.~\cite{bora}). In what
follows, we absorb the contribution of the $b_0$ term via
\beq
c_1 ' \, M_\pi^2= c_1 \, M^2 + 2b_0 M^4~.
\eeq
Explicit calculation using the
usual procedure of renormalization in CHPT~\cite{gl85}, leads to
the third order contribution to $\Sigma$,
\begin{eqnarray}\label{S3}
\hat{\Sigma}^{(3)}(\omega)
& = &
\Sigma^{(3)}_{\rm loop}(\omega) + \Sigma^{(3)}_{\rm div}(\omega)
\nonumber \\
& = &
\frac{3 g_A^2}{4 F_\pi^2} (M_\pi^2 - \omega^2)
\left( \frac{\omega}{8\pi^2} - \frac{1}{4\pi^2}
\sqrt{M_\pi^2 - \omega^2} \arccos \frac{-\omega}{M_\pi} \right) \nonumber\\
&& - \frac{3 g_A^2}{2 F_\pi^2}\, (3 M_\pi^2 - 2 \omega^2)
\frac{\omega}{16\pi^2} \ln \frac{M_\pi}{\lambda}
+ \omega^3 d^r_{24}(\lambda) - 8 M_\pi^2 \omega d^r_{28}(\lambda)~.
\end{eqnarray}
Similarly, the fourth order self--energy reads
\begin{eqnarray}\label{S4}
\hat{\Sigma}^{(4)}(\omega)
& = &
\Sigma^{(4)}_{\rm loop}(\omega) + \Sigma^{(4)}_{\rm div}(\omega)
\nonumber \\
& = &
-\frac{3 g_A^2}{8 m_N F_\pi^2}
\left(k^2 - 2\omega^2 + 8 m_N c_1 M_\pi^2 + M_\pi^2 \right)
\nonumber \\
& \times & \left(\frac{1}{8\pi^2}(\omega^2 + M_\pi^2)
-\frac{3\omega}{4\pi^2}\sqrt{M_\pi^2-\omega^2} \arccos\frac{-\omega}{M_\pi}
\right) 
+ \frac{3}{128 \pi^2 F_\pi^2} c_2 M_\pi^4 \nonumber \\
&-& \frac{9 g_A^2}{4 m_N F_\pi^2}
(k^2 - 2\omega^2 + 8 m_N c_1 M_\pi^2 + M_\pi^2)(M_\pi^2 - 2 \omega^2)
\frac{1}{16\pi^2} \ln \frac{M_\pi}{\lambda} \nonumber \\
&+& \left(
12 c_1 M_\pi^4 - 6 c_3  M_\pi^4
- \frac{3  M_\pi^4}{2} \biggl(c_2 - \frac{g_A^2}{8 m_N}\biggr)
\right)
\frac{1}{16\pi^2 F_\pi^2} \ln \frac{M_\pi}{\lambda}
\nonumber \\
&+& \frac{9 g_A^2 M_\pi^4}{256 \pi^2 m_N F_\pi^2} \ln \frac{M_\pi}{\lambda}
- 16 M_\pi^4 b^r_{21}(\lambda)
- 4 M_\pi^2 \omega^2 b^r_{160}(\lambda)
\nonumber \\
&-& 4 M_\pi^2 k^2 b^r_{161}(\lambda)
- \omega^4 b^r_{197}(\lambda)
- \omega^2 k^2 \left(b^r_{198}(\lambda)+b^r_{199}(\lambda)\right)
~, 
\end{eqnarray}
with $\lambda$ the scale of dimensional regularization.
All quantities have been set to their physical values since
the differences to the chiral limit values only appear at next
order. Of course,  eq.(\ref{S3}) agrees with the
third order self--energy expression given in~\cite{bkkm}.
The logarithms in eq.(\ref{S4}) stem from the self--energy graphs (fourth
last) and the tadpoles (third last line), respectively.
The $b^r_{21, \ldots , 199}$ are renormalized fourth order
LECs taken from table~1 of ref.\cite{mms4} (note that the fourth order
LECs are called $d_i$ in~\cite{mms4}. To avoid confusion with the
labelling in the FMS Lagrangian, we call them $b_i$ here).  
By a proper redefinition of the renormalized fourth order LECs, one can absorb
all the logarithmic terms,
\beq
b_i^r (\lambda ) = \frac{\beta_i}{16\pi^2} \left[ \bar{b}_i + \ln
  \frac{M}{\lambda} \right]~.
\eeq
Consequently, all $\ln (M_\pi / \lambda)$ terms vanish in eq.(\ref{S4}).
The corresponding $\bar{d}_i$ and $\bar{b}_i$ are all zero, as
detailed in~\cite{FMS}.
With a similar procedure for the $d_i$, one can also absorb all the
logarithms in eq.(\ref{S3}), for details see~\cite{FMS}.
{}From the pole position we read off the mass shift
\begin{eqnarray}
\delta m & = &
- \frac{2 m_0}{2 m_0 + \delta m} 4 c_1 ' \, M_\pi^2 
+ \frac{1}{2 m_0 + \delta m} 4 c_1 ' \, M_\pi^2 
\nonumber \\
& &
+ \frac{2 m_0  + vk - 4 c_1 ' \, M_\pi^2 
}{2 m_0 + \delta m}
  \left(\hat{\Sigma}^{(3)} (vk) + \hat{\Sigma}^{(4)}
(vk)\right)\biggl|_{\rm phys}~,
\end{eqnarray}
To proceed, we have to work out the value  $vk|_{\rm phys}$,
\beq
vk|_{\rm phys} =
\frac{p^2 - m_0^2 - k^2}{2 m_0} =
0 + \delta m - \frac{k^2}{2 m_N} + \ldots~.
\eeq
Let us comment on the $k^2$--dependent terms. The third order
self--energy  $\Sigma^{(3)}$ does not depend on $v k$, since
$v k = {\cal O}(q^2)$, i.e. such terms can not contribute
at third order. At fourth order, however, one gets a term of the
type $v k \,{\Sigma^{(3)\prime}}  (0)$.
The nucleon mass shift is, of course, not $k^2$--dependent.
At fourth order, all terms $\sim k^2$ cancel, 
as we will make explicit below. The fourth order mass shift reads
\beqa
\delta m^{(4)} & = &
 \frac{3 M_\pi^4 c_2}{128 \pi^2 F_\pi^2}
-\frac{3 g_A^2 M_\pi^4}{64 \pi^2 m_N F_\pi^2}
\nonumber \\
& &
-\frac{3 M_\pi^4}{32 \pi^2 F_\pi^2}
\left(
- 8 c_1 + c_2 + 4 c_3
\right)
\ln \frac{M_\pi}{\lambda}
-\frac{3 g_A^2 M_\pi^4}{32 \pi^2 m_N F_\pi^2}
\ln \frac{M_\pi}{\lambda}
\nonumber \\
& &
+ 16 M_\pi^4 \left(2 c_1 d^r_{28} - b^r_{21}\right)
+ 4 M_\pi^2 k^2 \left(\frac{d^r_{28}}{m_0} - b^r_{161}\right)
~,
\eeqa
with the pertinent $\beta$--functions,
$(4\pi F)^2 \,\beta (b^r_{161}) = -9g_A^2 / (16 m_0)$ and
$(4\pi F)^2 \,\beta (d^r_{28}) = -9g_A^2 / 16$. 
As promised, the $k^2$ terms add up to zero.
In that basis, the mass shift to fourth order takes the form
\beq
\delta m =
- 4 c_1 ' M_\pi^2
- \frac{3 g_A^2 M_\pi^3}{32 \pi F_\pi^2}
+ \frac{3 M_\pi^4 c_2}{128 \pi^2 F_\pi^2}
- \frac{3 g_A^2 M_\pi^4}{64 \pi^2 m_N F_\pi^2}
+ {\cal O} (q^5)~.
\eeq
Here, all masses and couplings are set to their physical values, the error
made by this procedure is of higher order. It is instructive to work
out these corrections numerically. Using the LECs as
determined in~\cite{bkmlec} (and ignoring the quark mass
renormalization of $c_1$, i.e. setting $c_1 ' = c_1$), we get
\beq
\delta m^{(4)} = ( 72.5 - 15.1 + 0.4 - 0.4 )~{\rm MeV}~,
\eeq
which shows that with the exception of the (undetermined)
${b}_0$--term (hidden in $c_1 '$), the fourth order corrections are tiny.
The general structure
of the fourth order contribution to the nucleon mass was already given
by Kallen~\cite{anne}, but we do not agree with some of her
coefficients.

We now come back to the  Z--factor. We have to generalize
eq.(\ref{Gn}) in that all possible loop effects have to be taken into
account. In a symbolic language, this reads
\beqa
G_n
& = &
( 1 + \underbrace{C_0^{-1} B_0}
) T_0^{-1} \; P_v^+ \, \hat{T} \, P_v^+
\; T_0^{-1} ( 1 + B'_0 C_0^{-1} ) \nonumber \\
& + &
( 1 + \underbrace{C_0^{-1} B_0 ) T_0^{-1}}
\; P_v^+ \, \hat{T} \, P_v^+
\; T_0^{-1} ( 1 + B'_0 C_0^{-1} ) \nonumber \\
& + &
( 1 + \underbrace{C_0^{-1} B_0 ) T_0^{-1} \; P_v^+ \,
  \hat{T}} \, P_v^+
\; T_0^{-1} ( 1 + B'_0 C_0^{-1} ) \nonumber \\
& + &
( 1 + \underbrace{C_0^{-1} B_0 ) T_0^{-1} \; P_v^+ \, \hat{T} \, P_v^+
\; T_0^{-1}} ( 1 + B'_0 C_0^{-1} ) \nonumber \\
& + & \ldots
\eeqa
where the 'underbrace' indicates that loops have to be constructed
from the pertinent operator structures involving $C_0^{-1}$ and
$B_0$. A closer look at these contributions reveals that only the
first two underbraced combinations of operators have the correct pole
structure, however, they only start to contribute at order $q^5$ (when
one counts the small momenta at the level of the Lagrangian).
This means that loops involving the heavy sources do indeed only
start to contribute at that order.
Consequently, for calculating the Z--factor to ${\cal O}(q^4)$
we have to work out (in analogy to eq.(\ref{Gnint}))
\beqa\label{Gnintl}
&& (\not\!p - m_N)(1 + C_0^{-1} B_0) [T_0 - \Sigma(\omega)] ^{-1}
\nonumber \\& = &
(\not\!p - m_N)( C_0 + B_0) [C_0 T_0 - C_0 \Sigma(\omega)]^{-1}
\nonumber \\
& = &
(\not\!p - m_N)
(\not\!p + m_0 - 4 c_1 M^2 - 8 b_0 M^4 + 2 vk P_v^-) 
[C_0 T_0 - C_0 \Sigma(\omega)]^{-1}~,
\eeqa
with $P_v^+  P_v^- = 0$. $Z_N$ is the value at the (physical) pole,
\beqa
Z_N^{-1} &  = &
(m_N + m_0 - 4 c_1 M^2 - 8 b_0 M^4)^{-1}
\frac{d}{d p} [C_0 T_0 - C_0 \Sigma(\omega)]|_{\rm phys}
\nonumber \\
& = & (m_N + m_0 - 4 c_1 M^2 - 8 b_0 M^4)^{-1}
\nonumber \\
& \times &
\left\{
2 m_N - \frac{m_N}{m_0} (\hat{\Sigma}^{(3)} (vk_{|{\rm phys}})
+\hat{\Sigma}^{(4)} (vk_{|{\rm phys}})) \right. \\
& &
\left.
- (2 m_0  + vk_{|{\rm phys}} - 4 c_1 M^2 - 8 b_0 M^4) \frac{m_N}{m_0}
  (\hat{\Sigma}^{(3) \prime}  (vk_{|{\rm phys}})
+\hat{\Sigma}^{(4) \prime} (vk_{|{\rm phys}}))
\right\}
\nonumber \\
& = &
\biggl( 2m_N + \frac{3 g_A^2 M_\pi^3}{32 \pi F_\pi^2}
+ {\cal O} (q^4) \biggr)^{-1}
\nonumber \\
& \times &
\left\{
2 m_N + \frac{3 g_A^2 M_\pi^3}{32 \pi F_\pi^2}
+ 2 m_N ( \frac{3 g_A^2 M_\pi^2}{32 \pi^2 F_\pi^2}
 - \frac{9 g_A^2 M_\pi^3}{64 \pi m_N F_\pi^2})
+ {\cal O} (q^4)
\right\}~,
\eeqa
where the prime denotes differentiation with respect to $p$ and we used
\beq
\frac{d}{dp} vk = \frac{m_N}{m_0} = 1 + {\cal O}(q^2)~.
\eeq
Finally, we arrive at
\begin{eqnarray}\label{ZDirac}
Z_N =
\Bigg\{
1 - \frac{3 g_A^2 M_\pi^2}{32 \pi^2 F_\pi^2}
  + \frac{9 g_A^2 M_\pi^3}{64 \pi m_N F_\pi^2}
\Bigg\}
+ {\cal O} (q^4)~.
\end{eqnarray}
Note that the first correction from two loops appears at order
$q^4$ since the Z--factor is the derivative of the self--energy
and thus one power of small momentum is absorbed in the definition
of $Z_N$. We remark that the Z--factor is momentum--independent and that, by
construction, it fulfills all four requirements spelled out in the
beginning of this section. We are now in the position to analyze
the various methods used so far in explicit HBCHPT calculations,
mostly based on a Pauli--spinor interpretation of the light fields.
To end this section, we note that in ref.\cite{eckmojwf}, the Dirac
interpretation was also used. In contrast to what is done here, field
redefinitions have been used to bring the effective Lagrangian into
a minimal form. We remark that the question
of the influence of such field redefinitions 
needs to be addressed in more detail. Such an investigation is underway
but goes beyond the scope of the present paper.

\section{Pauli spinor interpretation}
\label{sec:Pauli}
\def\theequation{\arabic{section}.\arabic{equation}}
\setcounter{equation}{0}

To clarify the approach  used by BKKM and also by FMS, we now interpret
the light fields $H$ as two--component Pauli spinors. Clearly, this
differs from the interpretation given in the preceeding section and the
crucial point will be the discussion of the contribution from
the heavy fields/sources, which in BKKM are hidden in the expansion
of the relativistic tree graphs and in FMS come in via the spinor
normalization factor.

To be more precise,
we reiterate the statement already made in sec.~\ref{sec:HBCHPT}
that in the rest--frame $v_\mu = (1,\vec{0}\,)$ a connection
between the relativistic and the Pauli--spinor interpretation
is given through the relation $P_v^+ \, u = {\cal N}\, (\chi, 0)^T$,
with $u$ ($\chi$) denoting the conventional four(two)--component
Dirac (Pauli) spinor. In light of these remarks, consider again
the generating functional. In the Pauli spinor interpreation,
the pion--nucleon action has no heavy sources, cf. eq.(\ref{Sprime}),
\beq
S_{\pi N}' = \int d^4x \, \bar{H}_v \bigl( A^{} +
B' \, C^{-1} B \, \bigr) H_v
+ \bar{H}_v \, R_v +  \bar{R}_v  \, H_v \, \, .
\eeq
Completing the square is achieved by setting
\beq
H_v' = H_v - T^{-1} \, R_v \, , \quad
T = A + B' \, C^{-1} \, B \,\,\, ,
\eeq
leading to
\beq
{\cal Z}  =  - \int d^4 x \bar{R}_v (A + B' C^{-1} B)^{-1} R_v~,
\eeq
which means that the other components do not play any role. The matrices
$A$,$B$ and $C$ are the standard ones when the relativistic Lagrangian
is turned into its heavy fermion form as explained in section~\ref{sec:HBCHPT}.
Consequently, the inverse propagator is given by
\beq
S^{-1} = A_0 + B_0' C_0^{-1} B_0 = T_0~,
\eeq
with $T_0$ given in eq.(\ref{T0}). Therefore, to this order one again
obtains the standard mass shift $\delta m = - 4 c_1 M^2$. The loop
corrections to this result will be discussed below. It is instructive
to first consider the Z--factor, which is defined as follows,
\begin{eqnarray}
Z_N^{-1}  & := &  \frac{d}{dp} S^{-1} \biggl|_{{\barr p} = m_N} \nonumber\\
& = &
(2 m_0 + vk - 4 c_1 M^2)^{-1}\biggl|_{\rm phys} \cdot
\frac{d}{dp} \left(p^2 - m_N^2 \right)\biggl|_{{\barr p} = m_N} \nonumber \\
& = & 2 m_N (m_N + vp)^{-1}~,
\end{eqnarray}
where ``phys'' means that the expression has to be evaluated at the
physical value of $vk$. Setting now $v=(1,0,0,0)$
(i.e. considering the rest--frame of the heavy nucleon) gives
\beq\label{ZPrest}
Z_N = \frac{m_N + E_p}{2 m_N} = {\cal N}^2~,
\eeq
with  ${\cal N}$ the normalization factor of the relativistic spinors,
already given in eq.(\ref{eq:N}). This underlines the observation made
by Fettes et al.~\cite{FMS}, namely that retaining the spinor normalization
allows to recover the  expanded full relativistic tree result directly
from the heavy nucleon approach (in that paper, pion--nucleon scattering
was investigated). Here, this observation is clearly more general.
In fact, the definition of the Z--factor given here fully justifies the
method used by BKKM (at least to one loop order $q^3$ and
$q^4$. Higher order calculations have not yet been attempted, with the
exception of~\cite{bora}~\cite{bkmff} and \cite{bimg}).

The self--energy calculation including the one loop effects
proceeds as outlined before. We only want
to stress that in the Pauli spinor interpretation one has the exact
same pole and thus the same mass shift as in the Dirac case discussed
in the previous section. The Z--factor is, however, different,
\begin{eqnarray}
Z^{-1}
& = &
(E_p + m_0 - 4 c_1 M^2 - 8 b_0 M^4)^{-1}
\nonumber \\
& \times &
\left\{
2 m_N - \frac{m_N}{m_0} (\hat{\Sigma}^{(3)} (vk_{|{\rm phys}})
+\hat{\Sigma}^{(4)} (vk_{|{\rm phys}})) \right. \\
& &
\left.
- (2 m_0  + vk_{|{\rm phys}} - 4 c_1 M^2 - 8 b_0 M^4) \frac{m_N}{m_0}
  (\hat{\Sigma}^{(3) \prime}  (vk_{|{\rm phys}})
+\hat{\Sigma}^{(4) \prime} (vk_{|{\rm phys}}))
\right\}
\nonumber \\
& = &
\biggl(E_p + m_N + \frac{3 g_A^2 M_\pi^3}{32 \pi F_\pi^2}
+ {\cal O} (q^4) \biggr)^{-1}   \nonumber \\
& \times &
\left\{
2 m_N + \frac{3 g_A^2 M_\pi^3}{32 \pi F_\pi^2}
+ 2 m_N ( \frac{3 g_A^2 M_\pi^2}{32 \pi^2 F_\pi^2}
 - \frac{9 g_A^2 M_\pi^3}{64 \pi m_N F_\pi^2})
+ {\cal O} (q^4)
\right\}~,
\end{eqnarray}
and thus
\begin{eqnarray} \label{ZPauli}
Z_N =
\frac{E_p + m_N}{2 m_N}
\Bigg\{
1 - \frac{3 g_A^2 M_\pi^2}{32 \pi^2 F_\pi^2}
  + \frac{9 g_A^2 M_\pi^3}{64 \pi m_N F_\pi^2}
\Bigg\}
+ {\cal O} (q^4)~.
\end{eqnarray}
Comparison with eq.(\ref{ZDirac}) gives
\beq
Z_N^{\rm Pauli} = {\cal N}^2 \, Z_N^{\rm Dirac}~.
\eeq
This shows that the factorization, which appeared somewhat {\it ad} {\it
  hoc} in eq.(\ref{ZFMS}), is exactly reproduced. It has its origin in
the precise matching of the  heavy baryon to the fully relativistic
theory. This is mirrored in the momentum--dependence of $Z_N$ in
eq.(\ref{ZPauli}) and it
justifies {\it a} {\it posteriori} the methods employed
by BKKM (at one loop order) and FMS. A precise statement about what
happens at higher orders can only be made when one performs explicit
calculations. This is not the aim of this paper.

Finally, we perform a calculation  based on the
Lagrangian given in~\cite{eckmoj} in the Pauli spinor interpretation
(we stress that this is not what was done in~\cite{eckmojwf}). 
Field redefinitions have been used to bring the Lagrangian in a
minimal form. This naturally changes the propagator, to third order 
it reads
\beq
S^{-1}  =
vk + \frac{k^2}{2 m_0} + \frac{4 a_3 M^2}{m_0} - \Sigma(vk)
=
\frac{p^2 - m_0^2}{2 m_0} + \frac{4 a_3 M^2}{m_0} - \Sigma(vk)~,
\eeq
with the LEC $a_3$ related to $c_1$ via $a_3 = m_N \, c_1$. Note that
we can not give the fourth order result since the corrections due to
the field redefinitions have not yet been worked out at this order.
The mass shift is, of course, identical to the one given so far,
but the Z--factor is different,
\beq\label{ZEM}
Z_N^{-1}  =
\frac{d}{dp} S^{-1} =
\frac{m_N}{m_0} \left( 1 - \Sigma^{(3)\prime} (0) \right)
+ {\cal O} (q^4)~,
\eeq
Note that in this formulation $Z_N$ is momentum--independent because
it has been worked out for Pauli spinors.  We finally remark that 
for HBCHPT, the Pauli interpretation can be considered the natural
framework.

\section{The charge form factor of the nucleon}
\label{sec:charge}
\def\theequation{\arabic{section}.\arabic{equation}}
\setcounter{equation}{0}

In this section, we explicitely calculate the isovector electric Sachs form
factor of the nucleon to third order in the chiral expansion.
This serves to illustrate the point that all
methods discussed so far lead to the same result if applied correctly.
It also shows in a very transparent way the differences of the various
frameworks in intermediate steps. A detailed discussion of the nucleon
form factors can be found in~\cite{BFHM}.

To be specific, consider the nucleon matrix element of the isovector component
of the
quark vector current $V_\mu^i =\bar q \gamma_\mu (\tau^i /2) q$ in the
Breit frame. It was already shown in ref.\cite{bkkm} that in HBCHPT
this is the natural frame (since the nucleon essentially behaves as
brick--wall with respect to the incomimg soft virtual
photon).  In the rest--frame $v_\mu = (1,\vec{0}\,)$,
the matrix--element of the isovector--vector current takes the form
(note that the superscript $'v'$ refers to the isovector current, not
the four--velocity)
\begin{equation}\label{Sachs}
\langle N(p')|V_{\mu}^i(0)|N(p)\rangle = 
\chi_2^\dagger
\left[
G^v_E(k^2) v_\mu + \frac{1}{m_N} G^v_M(k^2) [S_\mu,S_\nu] k^\nu
\right]
\chi_1
\times \eta^\dagger_2 \frac{\tau^i}{2} \eta_1~,
\end{equation}
where $\chi$ is a Pauli spinor with isospin component $\eta$ and
$k^2 = (p' - p)^2 <0$\footnote{The photon four--momentum
$k$ should not be confused with the same symbol previously denoting
the nucleons' small residual momentum.} is the invariant momentum transfer
squared.
${G}_E^v (k^2)$ and ${G}_M^v (k^2)$ are the isovector electric and magnetic
Sachs form factors. We remark that we can
replace the Pauli--Lubanski spin--vector $S_\mu$ by the Pauli spin
matrices, since $S_\mu = (\vec{\sigma},\vec{\sigma}\,)/2$
is restricted to the two upper components.
Obviously, we also need this matrix element
sandwiched between Dirac spinors, in which case it takes the form
\beq\label{SachsT}
\langle N(p')|V_{\mu}^i(0)|N(p)\rangle = 
\bar{u}(p') P_v^+  \left[
\tilde{G}^v_E(k^2) v_\mu + \frac{1}{m_N} \tilde{G}^v_M(k^2) [S_\mu,S_\nu] k^\nu
\right]
P^+_v u(p)
\times \eta^\dagger_2 \frac{\tau^i}{2} \eta_1~,
\eeq
where $u(p)$ is a Dirac spinor. Here,
$\tilde{G}_E^v (k^2)$ and $\tilde{G}_M^v (k^2)$ are related to the isovector
electric
and the isovector magnetic  Sachs form factor, via
\beq
\tilde{G}^v_{E,M} = \frac{1}{{\cal N}_1 {\cal N}_2} G^v_{E,M} =
\frac{2 m_N}{E + m_N} G^v_{E,M}~.
\eeq

We are now in the position to calculate these form factors. Note that
we calculate the objects $\tilde{G}_{E,M}^v$ within the Dirac spinor
framework whereas the $G_{E,M}^v$ follow in the Pauli spinor
interpretation. From here on, we concentrate on the charge (electric)
isovector form factor. The pertinent kinematics for the tree level
photon--nucleon coupling, cf. fig.\ref{fig1}, is
\beqa
&& p = m_0 \, v + p_1 \, , \quad
p' = m_0 \, v + p_2 \, , \quad
p_{1}^{\mu}=\left(E',-\frac{\vec{k}}{2}\right), \quad p_{2}^\mu=\left(E'
,+\frac{\vec{k}}{2}\right)~,  
\nonumber \\
&& k^\mu=\left(0,\vec{k} \, \right)\, , \quad
v\cdot k = 0 \,, \quad v \cdot p_1 = v\cdot p_2 = E' = E-m_0 = {\cal
  O} (q^2)~,
\eeqa
where $E'$ denotes the residual energy and $p_{1,2}$ the soft residual
momenta of the in-- and out--going nucleon, respectively.
We can rewrite the isovector Sachs form factors as
\FIGURE[l]{
\epsfxsize=3.5cm
\epsfbox{cnrt.epsi}
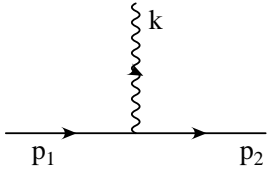
\caption{\label{fig1} Generic tree diagram for the
photon coupling to the nucleon.\bigskip}}
\noindent 
\beq
{G}_{E,M}^v (k^2) = Z_N \cdot  ( {\rm Born} \,\, {\rm
  terms} + {\rm  loops} )~,
\eeq
with the loop contribution being the same in all approaches. Their
explicit calculation is relegated to the end of this section.
For the moment, we concentrate on the Born (tree) terms and the
Z--factor, which are different in all schemes but when combined,
should lead to the same result, symbolically
\beq
G_E^{v, {\rm Born}} (k^2) 
= e~,
\eeq
as it follows from the relativistic calculation~\cite{gss}.
Note that from the Born terms
we only consider the ones with fixed coefficients in the $1/m_N$
expansion  since all others contribute to the anomalous magnetic
moment, the charge and magnetic radii and so on.

We now calculate explicitely the Born contribution
to $G_E^v (k^2)$ to third order in small momenta within the Pauli
spinor interpretation
(which is equivalent to FMS approach to this order).
We first collect all tree level terms
free of LECs that can contribute at third order  to the generic
diagram shown in fig.~\ref{fig1}. The pertinent operators and
respective Feynman rules (derived from the FMS Lagrangian)
for the photon--nucleon coupling are
\beqa
{\cal O}(1) &:& \quad i v \cdot D\to Q \, v
\cdot \epsilon~, \\
{\cal O}(2) &:& \quad \frac{1}{2m_0} (v \cdot D)^2 \to -\frac{1}{2m_0} Q
\, v \cdot \epsilon \, v\cdot (p_1 + p_2)~, \nonumber \\
&& -\frac{1}{2m_0} D^2 \to \frac{1}{2m_0} Q \,
 \epsilon \cdot (p_1 + p_2)~, \nonumber \\
&& -\frac{i}{4m_0} [S^\mu , S^\nu]F_{\mu\nu}^+ \to -\frac{1}{2m_0} 2Q
[S\cdot k , S \cdot \epsilon ]~,\\
{\cal O}(3) &:& -\frac{i}{4m_0^2} (v \cdot D)^3 \to \frac{1}{4m_0^2}
Q \, ((v\cdot p_1)^2 + (v\cdot p_2)^2 + v\cdot p_1
v\cdot p_2) v \cdot \epsilon~,\nonumber\\
&&\quad \frac{1}{8m_0^2} (iD^2\, v\cdot D + {\rm h.c.}) \to
-\frac{1}{8m_0^2} Q \,[ v\cdot \epsilon (p_1^2 + p_2^2) + v \cdot
 (p_1 + p_2) \epsilon \cdot  (p_1 + p_2) ]~, \nonumber\\
&& -\frac{1}{16m_0^2}\bigl( [S^\mu, S^\nu ] F_{\mu\nu}^+ v \cdot D +
{\rm h.c.} \bigr) \to  \frac{1}{4m_0^2}Q \, [S\cdot k , S \cdot \epsilon ]
v \cdot (p_1 + p_2 )~, \nonumber \\
&&-\frac{1}{8m_0^2} \bigl( [S^\mu, S^\nu ] F_{\mu\sigma}^+ v^\sigma D_\nu
+ {\rm h.c.} \bigr) \to \frac{1}{4m_0^2} Q \, \bigl( [ S\cdot k ,
S\cdot (p_1 + p_2) ] v \cdot \epsilon \nonumber \\
&& \qquad\qquad\qquad\qquad\qquad\qquad\qquad\qquad\quad\,\,\,
- [S\cdot \epsilon, S\cdot (p_1 + p_2) ] v \cdot k \bigr)~,\nonumber\\
&& -\frac{1}{16m_0^2} [D^\mu , F_{\mu\nu}^+ ] v^\nu \to \frac{1}{8m_0^2}
Q\, (k^2 v \cdot \epsilon - k\cdot \epsilon v \cdot k)~, \\
{\cal O}(4) &:& -\frac{1}{2m_0^3} S\cdot D \,(v \cdot D)^2 \, S\cdot D
\to -\frac{1}{2m_0^3}
Q \, [ (v\cdot p_1)^2 \,S\cdot \epsilon \, S\cdot p_1 
+ (v\cdot p_2)^2 \, S\cdot p_2 \, S\cdot \epsilon  \nonumber \\
&& \qquad\qquad\qquad\qquad\qquad\qquad\qquad\qquad\quad\,\,\,
+  v \cdot \epsilon \, S\cdot p_1 \,
S\cdot p_2 \, v\cdot (p_1 +p_2)]~,
\eeqa
with $\epsilon_\mu$ the polarization vector of the photon and
$Q = e\,(1+\tau^3)/2 =  e \, {\rm diag}(1,0)$ the nucleon charge matrix.
It is important to note that $\epsilon_\mu \sim {\cal O}(q)$, i.e.
a calculation with a term of dimension $q^n$ from the Lagrangian gives
$G_E^v$ at order $q^{n-1}$ (since the polarization vector is taken out).
With these rules, we can straightforwardly calculate the Born terms
corresponding to the electric Sachs form factor,
\beqa
{\rm Born}\,\, {\rm terms}  &=& e \, \biggl(1 + \frac{{3E'}^2}{4m_N^2}
- \frac{1}{8m_N^2} {\vec k \,}^2 -  \frac{1}{8m_N^2} (p_1^2 + p_2^2 +
4 {E'}^2) +\frac{1}{16m_N^3} \, E' \,{\vec k \,}^2  
\biggr)  \nonumber \\
&= & e\, \biggl( 1 - \frac{{\vec k\,}^2}{16m_N^2} 
\biggr) + {\cal O}(q^4)~,
\eeqa
since $p_1^2 = p_2^2 = {E'}^2 - {\vec k\,}^2/4$ and $E'$ is of chiral
order two. Expanding now the Born Z--factor, see eq.(\ref{ZPrest}),
to third order in momentum,
\beq
Z_N^{\rm Pauli} = {\cal N}^2 = 1 + \frac{\vec{k\,}^2}{16 m_N^2} + \ldots~,
\eeq
we finally get for the isovector electric Sachs form factor,
\beq
G_E^{v, \rm Born} (k^2) = {\cal N}^2 \cdot ( {\rm Born}\,\, {\rm terms})
= e\, \bigl(1 + {\cal O}(q^4) \bigr)~,
\eeq
i.e. the tree graphs give a constant form factor with the correct
charge to the order one is working.
\noindent
We now turn to the BKKM approach. It formally amounts to calculate
\beq
G_E^{v ,(\rm BKKM)} (k^2) =
Z_N^{\rm BKKM} \cdot ( \mbox{rel.  Born terms + loops} )~.
\eeq
As already stated, the calculation of the Born terms
 is trivial since the relativistic
Born graphs lead to a constant $G_E^v (k^2)$ and thus for this application,
the calculation is considerably simpler than using HBCHPT to work
out the tree graphs in the Pauli spinor interpretation as detailed
above.

Finally, we are left with the EM version. Here, the Born terms take yet another
form since they start from a different Lagrangian,
\beq
({\rm Born} \,\, {\rm terms})^{\rm EM} = e \, \biggl( 1 +
\frac{vp}{m_0} +\frac{k^2}{8 m_0} \biggr)~,
\eeq
with
\beq
vp = \delta m - \frac{k^2}{8 m_N^2} + \ldots~.
\eeq
Multiplying this with the Born term Z--factor as given in
eq.(\ref{ZEM}), one finds that due to the term $\sim  m_0/m_N$
the charge form factor is again constant and properly normalized.
We add a few remarks: First, the difference between the BKKM and FMS schemes
is $\delta G_E^v \sim (Z_N^{\rm BKKM} - Z_N^{\rm FMS}) \,$loops~$\sim {\cal
    O}(q^4)$, i.e. one would probably have to modify the BKKM scheme for the
wave function renormalization when it comes to two--loop (or higher
order)  calculations. We do not follow this subtlety here in detail,
but it should be kept in mind. Second, we turn our attention to the Dirac
spinor interpretation, which allows for comparison of all schemes.
The calculation of the Sachs form factors can be
summarized as follows:
\beqa
G^{v, \,\rm Dirac}_{E,M} (k^2)
& = &
\frac{E + m_N}{2 m_N} \tilde{G}^{v, \,\rm Dirac}_{E,M} (k^2) 
\nonumber \\
& = &
\frac{E + m_N}{2 m_N} Z^{\rm Dirac}_N 
\cdot (\mbox{Born terms + loops})
\nonumber \\
& = &
Z^{\rm Pauli}_N 
\cdot (\mbox{Born terms + loops})
\nonumber \\
& = &
G^{v, \,\rm Pauli}_{E,M} (k^2)
\eeqa
which means that the Pauli und the Dirac interpretation give exactly
the same result.

\medskip

In summary, there are many ways to arrive at the correct result,
however, given a certain scheme, one strictly has to stay within
its rules. It remains to be shown that the loops do indeed not
renormalize the charge, as it  follows from gauge invariance, but
only lead to a momentum dependence of the electric Sachs form factor.
Note that the calculation of the loop contribution is the same
in all three schemes, provided one has a prescription how to treat
the spin vector in $d$ dimensions. This calculation was first
performed in~\cite{bkkm} (for the isovector Dirac form factor).
In the appendix~\ref{app:nonren}, we show a different and somewhat
unusual way to arrive at the same result. Finally, we remark that the
considerations presented in this section can easily be extended to
fourth order. For the sake of brevity, we do not spell out the
pertinent details here.

\section{Summary and conclusions}
\label{sec:summ}
\def\theequation{\arabic{section}.\arabic{equation}}
\setcounter{equation}{0}

In this paper, we have considered the questions surrounding wave
function renormalization in heavy fermion effective field
theories. As an example, we have studied heavy baryon chiral
perturbation theory for two flavors and in the isospin limes.
The pertinent results of this investigation can be summarized as
follows:

\begin{enumerate}

\item[i)] The most natural and economic way of defining wave function
renormalization in heavy fermion effective field theories rests on
the interpretation of the light components of the heavy fields (like
e.g. the nucleons) as {\it Dirac} spinors. In that way, one can define
the Z--factor subject to four conditions as detailed in section~3. In
particular, the so--defined Z--factor is momentum--independent and
its tree graph contribution is equal to one. Furthermore, this prescription can
be extended to higher orders, i.e. beyond one loop, without problems.
Such an interpretation is mandated by the correct matching of the
heavy fermion EFT to its relativistic counterpart.

\item[ii)] All calculations performed so far in HBCHPT have been
done under the assumption that the light components of the heavy
fields are to be interpreted as {\it Pauli} spinors. We have shown
the equivalence between such an approach and the Dirac spinor
interpretation, provided one works in the rest--frame $v = (1,
\vec{0}\,)$ in the Pauli case.
This allows to {\it justify} a posteriori the methods
employed by BKKM ($1/m_N$ expansion of the relativistic tree
graphs independent of the spinor interpretation)
and by FMS (inclusion of the explicit four--dimensional spinor
normalization in the tree graphs calculated from HBCHPT). We also
pointed towards some potential complications which arise when one employs
field redefinitions.

\item[iii)] When applied correctly, all these different schemes
lead to the same physics. As an example, we have shown
how the tree result for the proton charge form factor,
$G_E^{\rm tree} (k^2) = e = {\rm const.}$, with $e$ the
proton charge and $k^2$ the photons' four--momentum squared, emerges in the
various calculational schemes. We have also discussed the
non--renormalization of the electric charge due to the loop graphs.

\item[iv)] Furthermore, we have studied the nucleon mass shift to fourth order
in the pion mass. Apart from an unspecified counter term contribution, which
formally amounts to a quark mass correction of a dimension two operator,
these corrections are tiny. 

\end{enumerate}

\noindent
We hope that with this paper, the long--standing question of wave
function renormalization in heavy fermion effective field theories
can finally be put to rest.

\acknowledgments
We have benefited from discussions with our colleagues
Gerhard Ecker, Thomas Hemmert, Norbert Kaiser and Martin 
Moj\v zi\v s. We are particularly
grateful to Gerhard Ecker for a critical reading of the manuscript.
One of us (S.S.) thanks Prof. H. Georgi and the Physics Department
at Harvard University for discussions and hospitality during a 
stay when part of this work was done. We are also grateful to
Judith McGovern and Mike Birse for a useful communication.

\bigskip

\appendix
\def\theequation{\Alph{section}.\arabic{equation}}
\setcounter{equation}{0}
\section{Precise definition of the Z--factor}
\label{app:Zdef}

We first repeat the standard steps to get from the generating
functional ${\cal Z} [j,\eta,\bar \eta]$ for bosonic fields (like e.g.
the pions) and fermions (like e.g. the nucleons)
coupled to $\bar \eta$ and $\eta$ in the presence of
some external sources $j$ (like e.g. photons or quark mass insertions)
to the S--matrix (to be precise, we consider processes with one
fermion line running through the pertinent Feynman graphs):
\begin{enumerate}

\item Differentiation with respect to the external sources (here:
      $j, \eta, \bar \eta$).

\item Multiplication with the physical propagator.

\item Multiplication with the spinors $u$ and $\bar{u}$.

\item Substitution of $\sqrt{Z} u $ by $ u_{phys}$ and similarly for $\bar{u}$.

\end{enumerate}
This leads to the form of eq.(\ref{Zone}). Thus, $Z_N$ is defined by
the product of the physical propagator and the two--point Greens
function. For our case, this leads to 
\begin{eqnarray}
Z_N 
& := & 
(\not\!p -m_N) 
(1 + C_0^{-1} B_0) P_v^+ T_0^{-1} P_v^+ (1 + B_0^\prime C_0^{-1})
\nonumber \\
& = & 
(1 + C_0^{-1} B_0) T_0^{-1} P_v^+ +
(1 + C_0^{-1} B_0) T_0^{-1} B_0^\prime C_0^{-1} P_v^- 
\nonumber \\
& = & 
P_v^+ +
\frac{E_p}{m_N} P_v^-
\nonumber \\
& = &
Z_N^+ P_v^+ + Z_N^- P_v^-~,
\end{eqnarray}
which means that the Z--factor appears as the sum of two terms, which
are proportional to the velocity projection operators. The part of
relevance for the present discussion is, of course, $Z_N^+$, since all
Greens functions corresponding to the positive velocity eigenstates
include a projector $P_v^+$. This is exactly what has been used
in eqs.(\ref{Gnint},\ref{Gnintl}).

\def\theequation{\Alph{section}.\arabic{equation}}
\setcounter{equation}{0}
\section{Non--renormalization of the electric charge due to loops}
\label{app:nonren}

\FIGURE[l]{
\epsfxsize=5.5cm
\epsfbox{cnr.epsi}
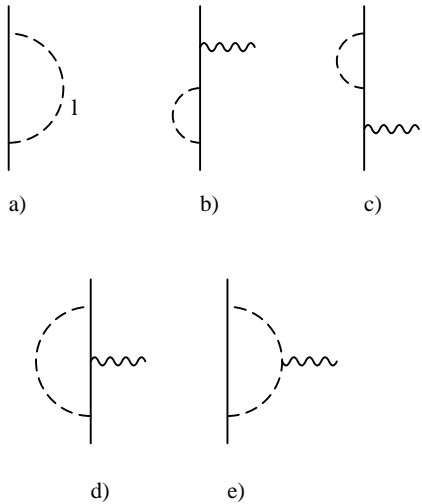
\caption{\label{fig2} a) Self-energy graph.
The solid and the dashed
line denote the nucleon and the pion, in order. b)-e) One loop graphs
contributing to the isovector charge form factor. The wiggly line
denotes the photon.\\$\,$
}}
\noindent
In this appendix, we explicitely show that the loops do indeed not
renormalize the charge, which follows from gauge invariance, but
only lead to a momentum dependence of the electric Sachs form factor.
Here, we entertain the possibility of a different prescription how to treat
the spin vector in $d$ dimensions. To be precise, as in ref.~\cite{bkkm} 
we consider  the 
Dirac form factor since at zero
momentum transfer, $F_1^v (0) =  G_E^v (0)$ because of the relation
$G_E (k^2) = F_1 (k^2) - (k^2/4m)F_2 (k^2)$. However, the
calculation shown it what follows differs from the
published version~\cite{bkkm} in that the spin matrices $S_\mu$ are not extended
to $d \neq 4$ dimension, but only the loop momenta $l$. 
Such a procedure can also be extended to higher orders, if needed.
For $S_\mu$, we consistently use the Pauli matrices, 
$S_\mu = (\vec{\sigma} \, , {\vec
  {\sigma}}\,)/2$. Therefore, $\vec \sigma \cdot \vec l$ picks out three
of the $d-1$ space--like components of $l$. Of course, the method used
in refs.\cite{bkkm}\cite{bkmrev} is correct, we only show this
alternative way to demonstrate that while the Z--factor depends on
such choices, physics does not (as long as one calculates correctly).
Clearly, the loop contribution to the Z--factor in this approach is
different to the one given in section~\ref{sec:Pauli}.
Consider first the self--energy graph in fig.2a. It gives (note that
to this order we can identify all quantities in the chiral limit with
their physical values, the error being of higher order)
\beq\label{CSE}
i \Sigma_{\rm loop} (\omega) = -i \frac{3g_A^2}{F_\pi^2} \, \int \frac{d^d
  l}{(2\pi )^d} \frac{ (-i)^2\,  S\cdot l S \cdot l}{(M_\pi^2 -
  l^2)(v\cdot l - \omega)} 
  = i \frac{9g_A^2}{4F_\pi^2} \, J_2
(\omega)~,
\eeq
using $S^2 = -3/4$, $S\cdot v = 0$, $\omega = v\cdot k$,
 and the loop function $J_2
(\omega)$ is given in app.~B of~\cite{bkmrev}. This leads to the
well--known mass shift (which we already derived in section~3, but
it is instructive to show that this somewhat unusual treatment of
the spin matrices indeed leads to the correct result)
\beq
\delta m^{\rm loop} = \Sigma_{\rm loop} (0)
= -\frac{3g_A^2M_\pi^3}{32\pi F_\pi^2}~,
\eeq
and the momentum--independent loop contribution to the Z--factor
\beq\label{CZN}
Z_N^{\rm loop}  = 1+ \Sigma_{\rm loop} ' (0) = \frac{9g_A^2M_\pi^2}{32\pi^2
  F_\pi^2}\ln\frac{M_\pi}{\lambda}\,\,\, {\rm mod}\, L(\lambda ) \,\, ,
\eeq
where we are not concerned with the infinite piece $\sim L$ in what
follows since it cancels as in described in~\cite{bkkm} and the prime
denotes differentiation with respect to $\omega$. To arrive at
this result, we have used $J_2 ' (0) = - \Delta_\pi = -
(M_\pi^2/8\pi^2 ) \ln (M_\pi / \lambda )$ (dropping again the
term $\sim L$). Note the difference
to $Z_N^{\rm loop}$ in~\cite{bkkm}, where one has $(3\ln(M_\pi/\lambda) +1)$
instead of $3\ln(M_\pi / \lambda )$
(in a short-hand notation). In particular, in this formulation only
the logarithmic terms survive. We do not absorb these in the LECs as
done before. We now consider the  charge form factor
at $k^2 = 0$. The graphs~1b,c contain $Z_N^{\rm loop}$, 
for completeness we give the result for the isoscalar as well 
as the isovector form factor
\beq
{\rm Amp} (1b+1c) = \frac{1+\tau_3}{2} (Z_N^{\rm loop} -1 ) = -(1+\tau_3 )
\frac{9g_A^2M_\pi^2}{64\pi^2  F_\pi^2}\ln\frac{M_\pi}{\lambda}~.
\eeq
The two one--loop diagrams 1d and 1e give
\beqa
{\rm Amp} (1d) &=& \frac{3-\tau_3}{2} \biggl(-\frac{3g_A^2}{4F_\pi^2}\biggr) \,
J_2 ' (0) = (3-\tau_3 ) \frac{3g_A^2M_\pi^2}{64\pi^2
F_\pi^2}\ln\frac{M_\pi}{\lambda}~, \\
{\rm Amp} (1e) &=& - \tau_3 \, \frac{g_A^2}{F_\pi^2} \, \gamma_8 (0) =
\tau_3 \, \frac{3g_A^2M_\pi^2}{64\pi^2 F_\pi^2}\ln\frac{M_\pi}{\lambda}~,
\eeqa
using $\gamma_8 (0) = -\Delta_\pi / 2$ (see app.~B of~\cite{bkmrev}).
Adding up pieces, we end up with
\beq
{\rm Amp} (1b+1c+1d+1e) = \frac{3g_A^2M_\pi^2}{64\pi^2
F_\pi^2}\ln\frac{M_\pi}{\lambda}\, \bigl[ -3-3\tau_3 + 3 - \tau_3 +
4\tau_3 \bigr] = 0~,
\eeq
which is nothing but the anticipated result. The terms $\sim L$,
which we did not give, also cancel. But this is a different statement,
since the cancellation of the infinities only means that one has
renormalized properly, and thus it does not constrain the finite
pieces we have discussed here. Finally, we remark that the
considerations presented in this section can easily be extended to
the next order. For the sake of brevity, we do not spell out the
pertinent details here.


\end{document}